\documentstyle[12pt]{article}        
\begin{document}
\def\ie{\hbox{\it i.e.}} \def\etc{\hbox{\it etc.}}
\def\eg{\hbox{\it e.g.}} \def\cf{\hbox{\it cf.}}
\def\etal{\hbox{\it et al.}}
\def\dash{\hbox{---}}
\def\cok{\mathop{\rm cok}}
\def\tr{\mathop{\rm tr}}
\def\Tr{\mathop{\rm Tr}}
\def\Im{\mathop{\rm Im}}
\def\Re{\mathop{\rm Re}}
\def\bR{\mathop{\bf R}}
\def\bC{\mathop{\bf C}}
\def\lie{\hbox{\it \$}} 
\def\partder#1#2{{\partial #1\over\partial #2}}
\def\secder#1#2#3{{\partial^2 #1\over\partial #2 \partial #3}}
\def\bra#1{\left\langle #1\right|}
\def\ket#1{\left| #1\right\rangle}
\def\VEV#1{\left\langle #1\right\rangle}
\let\vev\VEV
\def\gdot#1{\rlap{$#1$}/}
\def\abs#1{\left| #1\right|}
\def\pri#1{#1^\prime}
\def\ltap{\raisebox{-.4ex}{\rlap{$\sim$}} \raisebox{.4ex}{$<$}}
\def\gtap{\raisebox{-.4ex}{\rlap{$\sim$}} \raisebox{.4ex}{$>$}}


\def\lesssim{\mathrel{\mathpalette\vereq<}}
\def\gtrsim{\mathrel{\mathpalette\vereq>}}
\makeatletter
\def\vereq#1#2{\lower3pt\vbox{\baselineskip1.5pt \lineskip1.5pt
\ialign{$\m@th#1\hfill##\hfil$\crcr#2\crcr\sim\crcr}}}
\makeatother

\newcommand{\rem}[1]{{\bf #1}}

\renewcommand{\thefootnote}{\fnsymbol{footnote}}
\setcounter{footnote}{0}
\begin{titlepage}
\begin{center}

\hfill OUTP-9901P\\ 
\hfill UCB-PTH-98/57\\
\hfill LBNL-42572\\
\hfill SNS-PH/98-25\\
\hfill hep-ph/9901228\\
\hfill \today\\

\vskip .5in

{\Large \bf
Nearly Degenerate Neutrinos \\ and Broken Flavour Symmetry
\footnote
{This work was supported in part by the U.S.
Department of Energy under Contracts DE-AC03-76SF00098, in part by the
National Science Foundation under grant PHY-95-14797 and in part by the TMR
Network under the EEC Contract No. ERBFMRX-CT960090.}
}

\vskip .50in

Riccardo Barbieri$^1$, Lawrence J. Hall$^2$, G.L. Kane$^3$ and Graham G.
Ross$^4$

\vskip 0.05in

{\em $^1$ Scuola Normale Superiore, and\\
INFN, sezione di Pisa, I-56126 Pisa, Italia\\

$^2$ Department of Physics and Lawrence Berkeley National Laboratory\\
University of California, Berkeley, CA 94720

$^3$ Department of Physics, University of Michigan, Ann Arbor, MI48109\\

$^4$ Theoretical Physics, Oxford University, 1 Keble Rd., Oxford, OX13RH,
U.K.}

\vskip 0.05in

\end{center}

\vskip .5in

\begin{abstract}
Theories with non-Abelian flavour symmetries lead at zeroth order
to neutrino degeneracy and massless charged fermions. The flavour
symmetry is spontaneously broken sequentially to give hierarchies
$\Delta m_{atm}^2 \gg \Delta m_{\odot}^2$ and $m_\tau \gg m_\mu \gg
m_e$, and a misalignment of the vacuum between neutrino and charged
sectors gives large $\theta_{atm}$. Explicit models are given which
determine $\theta_{atm} = 45^\circ$.
A similar construction gives vacuum misalignment for the lighter generations
and gives a
vanishing $\beta \beta_{0 \nu}$ rate, so that there is no laboratory
constraint on the amount of neutrino hot dark matter in the universe,
and the solar mixing angle is also maximal, $\theta_\odot = 45^\circ$.

\end{abstract}
\end{titlepage}

\renewcommand{\thepage}{\arabic{page}}
\setcounter{page}{1}
\renewcommand{\thefootnote}{\arabic{footnote}}
\setcounter{footnote}{0}

\section{Introduction}

Measurements of both solar and atmospheric neutrino fluxes provide
evidence for neutrino oscillations. With three neutrinos, this implies
that there is negligible neutrino hot dark matter in the universe unless
the three neutrinos are approximately degenerate in mass. In this letter
we construct theories with approximately degenerate neutrinos, consistent
with the atmospheric and solar data, in which the lepton masses and
mixings are governed by spontaneously broken flavour symmetries.

The Super-Kamiokande collaboration has measured the magnitude and angular
distribution of the $\nu_\mu$ flux originating from cosmic ray induced
atmospheric showers \cite{SK}.
They interpret the data in terms of large angle
($\theta > 32^\circ$) neutrino oscillations, with $\nu_\mu$ disappearing to
$\nu_\tau$ or a singlet neutrino with $\Delta m^2_{atm}$ close to
$10^{-3} \mbox{eV}^2$. Five independent solar neutrino experiments,
using three detection methods, have measured solar neutrino fluxes which
differ significantly from expectations. The data is consistent with
$\nu_e$ disappearance neutrino oscillations, occuring either inside the sun,
with $\Delta m^2_\odot$ of order $10^{-5} \mbox{eV}^2$, or between the sun
and the earth, with $\Delta m^2_\odot$ of order $10^{-10} \mbox{eV}^2$.
The combination of data on atmospheric and solar neutrino fluxes
therefore implies a hierarchy of neutrino mass splittings:
$\Delta m^2_{atm} \gg \Delta m^2_\odot$ \footnote{A problem in one of the
solar neutrino experiments or in the Standard Solar Model
could, however, allow comparable mass differences}.

In this letter we consider theories with three neutrinos. Ignoring the
small contribution to the neutrino mass matrix which gives $\Delta
m^2_\odot$, there are three possible forms for the neutrino mass
eigenvalues:
\begin{eqnarray}
\mbox{``Hierarchical''} \hspace{1in} \overline{m}_\nu &=& m_{atm}
\pmatrix{0&&\cr &0&\cr &&1} \hspace{1in} \label{eq:H} \\
\mbox{``Pseudo-Dirac''} \hspace{1in} \overline{m}_\nu &=& m_{atm}
\pmatrix{1&&\cr &1&\cr &&\alpha} \hspace{1in} \label{eq:PD} \\
\mbox{``Degenerate''} \hspace{1in} \overline{m}_\nu &=& m_{atm}
\pmatrix{0&&\cr &0&\cr &&1} + M \pmatrix{1&&\cr &1&\cr &&1}
\label{eq:D}
\end{eqnarray}
where $m_{atm}$ is approximately $0.03$ eV, the scale of the atmospheric
oscillations. The real parameter $\alpha$ is either of order unity (but
not very close to unity) or zero, while the mass scale $M$ is much larger
than $m_{atm}$. We have chosen to order the eigenvalues so that
$\Delta m^2_{atm} = \Delta m^2_{32}$, while $\Delta m^2_\odot = \Delta
m^2_{21}$
vanishes until perturbations much less than $m_{atm}$ are added.
An important implication of the Super-Kamiokande atmospheric data is that
the mixing $\theta_{\mu \tau}$ is large. It is remarkable that this large
mixing occurs between states with a hierarchy of $\Delta m^2$, and this
places
considerable constraints on model building.

What lies behind this pattern of neutrino masses and mixings? An
attractive possibility is that a broken flavour symmetry leads to the
leading order masses of (\ref{eq:H}), (\ref{eq:PD}) or (\ref{eq:D}), to
a large $\theta_{atm}$, and to acceptable values for $\theta_\odot$ and
$\Delta m^2_\odot$. It is simple to construct flavour symmetries which
lead to (\ref{eq:H}) or (\ref{eq:PD}) with large (although not necessarily
maximal) $\theta_{atm}$ \cite{large}. For
example, the hierarchical case results from integrating out a heavy Majorana
right-handed neutrino which has comparable complings to $\nu_\mu$ and
$\nu_\tau$, and the pseudo-Dirac case when the heavy state is Dirac, with
one component coupling to the $\nu_{\mu,\tau}$ combination
and the other to $\nu_e$.\footnote{The conventional paradigm for models
with flavour symmetries is the hierarchical case with hierarchically small
mixing angles, typically given by $\theta_{ij} \approx (m_i/m_j)^{{1
\over 2}}$. If the neutrino mass hierarchy is moderate, and if the
charged and neutral contributions to $\theta_{atm}$ add, this
conventional approach is not excluded by the data
\cite{smallangle}.} However, in both
hierarchical and pseudo-Dirac cases, the neutrino masses have upper
bounds of $(\Delta m^2_{atm})^{{1 \over 2}}$. In these schemes the sum of
the
neutrino masses is also bounded, $\Sigma_i m_{\nu i} \leq 0.1$ eV, implying
that neutrino hot dark matter has too low an abundance to be relevant for
any cosmological or astrophysical observation \cite{hdm}.

By contrast, it is more difficult to construct
theories with flavour symmetries for
the degenerate case \cite{degenerate}, where $\Sigma_i m_{\nu i} = 3M$,
which
are therefore unconstrained by any oscillation data. While non-Abelian
symmetries
can clearly obtain the degeneracy of (\ref{eq:D}) at zeroth order, the
difficulty is in obtaining the desired lepton mass hierarchies and mixing
angles, which requires flavour symmetry breaking vevs pointing in very
different directions in group space. We propose a solution to this vacuum
misalignment problem, and use it to construct a variety of models, some
of which predict $\theta_{atm} = 45^\circ$.
We also construct a model with bimaximal mixing \cite{bimax} 
having $\theta_{atm} = 45^\circ$ and $\theta_{12}= 45^\circ$ \cite{gg}.

\section{Texture Analysis}

What are the possible textures for the degenerate case in the flavour
basis? These textures will provide the starting point for constructing
theories with flavour symmetries.
In passing from flavour basis to mass basis, the relative
transformations of $e_L$ and $\nu_L$ gives the leptonic mixing matrix
$V$ \cite{mns}. Defining $V$ by the charged current in the mass basis,
$\overline{e} V \nu$, we choose to parameterize $V$ in the form
\begin{equation}
V = R(\theta_{23}) R(\theta_{13}) R(\theta_{12})
\label{eq:V}
\end{equation}
where $R(\theta_{ij})$ represents a rotation in the $ij$ plane by
angle $\theta_{ij}$, and diagonal phase matrices are left implicit.
The angle $\theta_{23}$ is necessarily large as it is
$\theta_{atm}$. In contrast, the Super-Kamiokande data constrains
$\theta_{13} \leq 20^\circ$ \cite{less20}, and if $\Delta m^2_{atm} > 2
\times
10^{-3} \mbox{eV}^2$, then the CHOOZ data requires $\theta_{13}
\leq 13^\circ$ \cite{CHOOZ}. For small angle MSW oscillations in the sun
\cite{msw},
$\theta_{12} \approx 0.05$, while other descriptions of the solar
fluxes require large values for $\theta_{12}$ \cite{solar}.

Which textures give such a $V$ together with the degenerate mass
eigenvalues of eqn. (\ref{eq:D})?
In searching for textures, we require that in the flavour basis
any two non-zero entries are either independent or equal up to a phase,
as could follow simply from flavour symmetries.
This allows just
three possible textures for $m_\nu$ at leading order

\begin{eqnarray}
``A'' \hspace{0.25in} m_\nu & = & M \pmatrix{1 & 0 & 0 \cr
0&1&0 \cr 0&0&1} + m_{atm} \pmatrix{0 & 0 & 0 \cr 0&0&0 \cr 0&0&1}
\label{eq:A}\\
``B'' \hspace{0.25in} m_\nu & = & M \pmatrix{0 & 1 & 0 \cr
1&0&0 \cr 0&0&1} + m_{atm} \pmatrix{0 & 0 & 0 \cr 0&0&0 \cr 0&0&1}
\label{eq:B}\\
``C'' \hspace{0.25in} m_\nu & = & M \pmatrix{1 & 0 & 0 \cr
0&0&1 \cr 0&1&0} + m_{atm} \pmatrix{0 & 0 & 0 \cr 0&1&-1 \cr 0&-1&1}
\label{eq:C}
\end{eqnarray}
Alternatives for the perturbations proportonal to $m_{atm}$ are
possible. Each of these textures will have to be coupled
to corresponding suitable textures for the charged lepton
mass matrix $m_E$, defined by $\overline{e_L} m_E e_R$. For example, in
cases
(A) and (B), the big $\theta_{23}$ rotation angle will have
to come from the diagonalization of $m_E$.

To what degree are the
three textures A,B and C the same physics written in different bases, and
to what extent can they describe different physics?
Any theory with degenerate neutrinos
can be written in a texture A form, a texture B form or a texture C
form, by using an appropriate choice of basis.
However, for certain cases, the physics may be more transparent
in one basis than in another, as illustrated later.

\section{A Misalignment Mechanism}

The near degeneracy of the three neutrinos requires a non-Abelian flavour
symmetry, which we take to be $SO(3)$, with the three lepton doublets,
$l$, transforming as a triplet. This is for simplicity -- many
discrete groups, such as a twisted product of two $Z_2$s would also give
zeroth order neutrino degeneracy. We expect the $SO(3)$ theories
discussed below to have analogues with discrete non-Abelian symmetries
\footnote{$SO(3)$ has been invoked recently in connection with
quasi-degenerate neutrinos also in Refs \cite{SO3}.}.

We work in a supersymmetric theory and
introduce a set of ``flavon'' chiral superfields which
spontaneously break SO(3). For now we just assign the desired vevs to
these fields; later we construct potentials which force these orientations.
Also, for simplicity we assume one set of flavon fields,
$\chi$, couple to operators which give neutrino masses, and another set,
$\phi$, to operators for charged lepton masses. We label fields according
to the direction of the vev, e.g. $\phi_3 = (0,0,v)$.
For example, texture A, with
\begin{equation}
m_E = \pmatrix{0 & 0 & 0 \cr 0&\delta_2&D_2 \cr 0&\delta_3&D_3}\equiv
m_{II},
\hspace{0.5in}
\end{equation}
results from the superpotential
\begin{equation}
W = (l \cdot l)hh +
(l \cdot \chi_3)^2 hh + (l \cdot \phi_3) \tau h +
(l \cdot \phi_2) \tau h + (l \cdot \phi_3) \xi_\mu \mu h
+ (l \cdot \phi_2) \xi_\mu \mu h
\label{eq:Aops}
\end{equation}
where the coefficient of each operator is understood to
be an order unity coupling multiplied by the appropriate inverse power of
the large flavour mass scale $M_f$.
The lepton doublet $l$ and the $\phi, \chi$ flavons are all
$SO(3)$ triplets, while the right-handed
charged leptons ($e, \mu, \tau$) and the Higgs doublets, $h$, are $SO(3)$
singlets. The electron mass is neglected.
The form of eqn. (\ref{eq:Aops}) may be guaranteed by additional
Abelian flavour symmetries; in the limit where these symmetries are
exact, the only charged
lepton to acquire a mass is the $\tau$. These
symmetries are broken by vevs of flavons $\xi_{e, \mu}$, which are
$SO(3)$ and standard model singlet fields. The hierarchy of charged
fermion masses is then generated by $\vev{\xi_{e, \mu}} / M_f$.
The ratios $\vev{\phi_{2,3}} / M_f$ and $\vev{\chi}/M_f$
generate small dimensionless $SO(3)$ symmetry breaking parameters.
The first term of (\ref{eq:Aops}) generates an $SO(3)$ invariant mass
for the neutrinos corresponding to the first term in (\ref{eq:A}). The
second term gives the second term of (\ref{eq:A}) with $m_{atm}/M =
\vev{\chi_3}^2 / M_f^2$. The remaining terms generate the charged
lepton mass matrices. Note that the charged fermion masses
vanish in the $SO(3)$ symmetric
limit --- this is the way we reconcile the near degeneracy of the
neutrino spectrum with the hierarchical charged lepton sector. This is
viable because, although the leading neutrino masses are $SO(3)$ invariant,
they are second order $SU(2)$ violating and are suppressed
relative to the electroweak scale $\vev{h}$ by $\vev{h} / M_f$, where
$M_f$ may be very large, of order the unification or Planck scale.
On the other hand the charged lepton masses, although arising only
via $SO(3)$ breaking, are only first order in $SU(2)$ breaking. Hence
their suppression relative to $\vev{h}$ is of order
$\vev{\phi_i} / M_f$. Since $\phi_i$
are $SU(2)$ singlets, they may have vevs much larger than $h$:
the charged leptons can indeed be much heavier than the neutrinos.

In this example we see that the origin of large $\theta_{atm}$ is due to
the misalignment of the $\phi$ vev directions relative to that of the
$\chi$ vev. This is generic.
In theories with flavour symmetries, large $\theta_{atm}$ can only arise
because of a misalignment of flavons in charged and neutral
sectors. To obtain $\theta_{atm} = 45^\circ$, as preferred by the
atmospheric data, requires however a very precise misalignment, which can
occur as follows. In a basis
where the $\chi$ vev is in the direction $(0,0,1)$, there should be a
single $\phi$ field coupling to $\tau$ which has a vev in the
direction $(0,1,1)$, where an independent phase for each entry is
understood.
As we shall now discuss, in theories based on $SO(3)$,
such an alignment occurs very easily,
and hence should be viewed as a typical expectation, and certainly
not as a fine tuning.

Consider any 2 dimensional subspace within the $l$ triplet, and label
the resulting 2-component vector of $SO(2)$ as $\ell = (\ell_1, \ell_2)$.
At zeroth order in SO(2) breaking only the neutrinos of $\ell$ acquire a
mass, and they are degenerate from $\ell \cdot \ell hh$. Introduce a flavon
doublet $\chi = (\chi_1, \chi_2)$ which acquires a vev to break $SO(2)$.
If this field were real, then one could do an $SO(2)$ rotation to set
$\vev{\chi_2} =0$. However, in supersymmetric theories
$\chi$ is complex and
a general vev has the form $\vev{\chi_i} = a_i + ib_i$.
Only one of these four real parameters can be set to zero using $SO(2)$
rotations. Hence the scalar potential can determine a variety of
interesting alignments. There are two alignments which are easily
produced and are very useful in constructing theories:
\begin{equation}
\mbox{``SO(2)'' Alignment:} \hspace{0.4in} W = X(\chi^2 - M^2);
\hspace{0.4in}
m_\chi^2 >0; \hspace{0.4in} \vev{\chi} = M(0,1).
\label{eq:so2al}
\end{equation}
The parameter $M$, which could result from the vev of some $SO(2)$
singlet, can be taken real and positive by a phase choice
for the fields. Writing $\vev{\chi_i} = a_i + i b_i$, with
$a_i$ and $b_i$ real, an $SO(2)$ rotation can always be done to set $a_1 =
0$.
The driver field $X$ forces $\vev{\chi^2} = M^2$, giving $b_2 = 0$
and $a_2^2 = b_1^2 + M^2$ with $b_1$ undetermined.
The potential term which aligns the directon of the $\chi$ vev is the
positive soft mass squared $m_{\chi^2} \chi^* \chi$, which
sets $b_1 = 0$.

The second example is:
\begin{equation}
\mbox{``U(1)'' Alignment:} \hspace{0.4in} W = X\varphi^2; \hspace{0.3in}
m_\varphi^2 <0; \hspace{0.3in} \vev{\varphi} = V(1,i) \mbox{ or } V(1,-i).
\label{eq:u1al}
\end{equation}
It is now the negative soft mass squared which forces a magnitude
$\sqrt{2}|V|$
for the vev.
Using $SO(2)$ freedom to set $a_2 = 0$, $|F_X|^2$
provides the aligning potential
and requires $\vev{\varphi}^2 = a_1^2 + 2i a_1 b_1 - b_1^2 - b_2^2 = 0$,
implying $b_1 = 0$ and $b_2 = \pm i a_1$. The $U(1)$ alignment leaves
a discrete 2-fold degeneracy. In fact, the vevs $V(1, \pm i)$
do not require any particular choice of $SO(2)$ basis: performing $SO(2)$
transformation by angle $\theta$ on them just changes the phase of $V$ by
$\pm \theta$.
The phases in $\vev{\varphi}$ are unimportant in determining the values of
the neutrino mixing angles, so that the relative orientation
of the vevs of (\ref{eq:so2al}) and (\ref{eq:u1al}) corresponds
to $45^\circ$ mixing.

The vev of the $SO(2)$ alignment, (\ref{eq:so2al}), picks out the original
$SO(2)$ basis; however, the vev of the $U(1)$ alignment,
(\ref{eq:u1al}), picks out a new basis $(\varphi_+, \varphi_-)$,
where $\varphi_\pm = (\varphi_1 \pm i
\varphi_2)/ \sqrt{2}$. If $\vev{(\varphi_1, \varphi_2)} \propto (1,i)$, then
$\vev{(\varphi_-, \varphi_+)} \propto (1,0)$. An important feature of
the $U(1)$ basis is that the $SO(2)$ invariant $\varphi_1^2 +
\varphi_2^2$ has the form $2 \varphi_+ \varphi_-$. In the SO(3)
theory, we usually think of $(l \cdot l) hh$ as
giving the unit matrix for neutrino masses as in texture A. However, if
we use the $U(1)$ basis for the 12 subspace, this operator actually gives
the leading term in texture B, whereas if we use the $U(1)$ basis in the
23 subspace we get the leading term in texture C.

\section{The Neutrinoless Double Beta Decay Constraint}

Searches for neutrinoless double beta decay, $\beta \beta_{0 \nu}$, place
a limit $m_{\nu ee} < 0.5$ eV \cite{2beta}.
Consider texture A with $m_E = m_{II}$, so that the electron is
dominantly in $l_1$. The $\beta \beta_{0 \nu}$
limit implies $\Sigma_i m_{\nu i} < 1.5$ eV,
and therefore places a constraint on the amount of neutrino hot dark
matter in the universe
\begin{equation}
\Omega_\nu (l_1 \simeq e) < 0.05 \left({0.5 \over h} \right)^2.
\label{eq:doublebeta}
\end{equation}
While values of $\Omega_\nu$ which satisfy this constraint
can be of cosmological interest, it is also important to know whether
this bound can be violated.

The bound is not greatly altered if texture A is taken with
\begin{equation}
m_E = \pmatrix{0 & \delta_1 & D_1 \cr 0&\delta_2&D_2 \cr
0&\delta_3&D_3}\equiv m_{III},
\end{equation}
for generic values of $D_1, D_2$ and $D_3$. However, there
is a unique situation where the $\beta \beta_{0 \nu}$ bound on the amount
of neutrino hot dark matter is evaded.
It is convenient to discuss this special case in the basis in which it
appears as texture B with $m_E = m_{II}$.
To the order which we work, the electron mass
eigenstate is then in the doublet $l_- = (l_1 - i l_2)/ \sqrt{2}$,
where we label the basis by $(-,+,3)$
and, since there is no neutrino mass
term $l_- l_- hh$, the rate for neutrinoless double beta decay vanishes.
This important result is not transparent
when the theory is described by texture A.
In this case $m_E = m_{III}(\delta_1 = i\delta_2, D_1 = iD_2)$
and the electron is in a linear combination of $l_1$ and $l_2$. There are
contributions to $\beta \beta_{0 \nu}$ from both $l_1 l_1 hh$ and
$l_2 l_2 hh$ operators, and these contributions cancel.

As an illustration of the utility of the $U(1)$ vev
alignment, this theory with vanishing $\beta \beta_{0\nu}$ rate is
described by the superpotential
\begin{equation}
W = (l \cdot l)hh + (l \cdot \chi_3)^2 hh + (l \cdot \phi_3) \tau h +
(l \cdot \phi_-) \tau h + (l \cdot \phi_3) \xi_\mu \mu h +
(l \cdot \phi_-) \xi_\mu \mu h.
\label{eq:hdmops}
\end{equation}
Comparing with the theory for texture A with $m_E = m_{II}$, described by
(\ref{eq:Aops}), the only change is the replacement of a vev in the 2
direction with one in the $-$ direction.

In theories of this sort, it is likely that a higher order
contribution to $\beta \beta_{0 \nu}$ will result when perturbations
are added for $m_e$ and $\Delta m^2_\odot$. For example, if the
electron mass results from mixing with the second generation by an
angle $\theta \simeq (m_e / m_\mu)^{{1 \over 2}}$, then $\beta
\beta_{0 \nu}$ is reintroduced. However, the resulting limit on
$\Omega_\nu$ is weaker than (\ref{eq:doublebeta}) by about an order of
magnitude, corresponding to this mixing angle. Large values of
$\Omega_\nu$ in such theories could be probed by further searches for
neutrinoless double beta decay.

\section{Models For Large $\theta_{atm}$}

Along the lines described above,
we first construct a model with large, but undetermined, $\theta_{atm}$,
which explicitly gives both the Yukawa couplings and the
orientation of the flavon vevs.
Introduce two SO(3) triplet flavons, carrying discrete symmetry charges
so that one, $\chi$, gives only neutrino masses, while the other, $\phi$,
gives only charged lepton masses:
\begin{equation}
W = (l \cdot l) hh + (l \cdot \chi)^2 hh + (l \cdot \phi) \tau h.
\label{eq:massops}
\end{equation}
Suppose that both flavons are forced to acquire vevs using the
``$SO(2)$'' alignment of (\ref{eq:so2al}):
\begin{equation}
W = X( \chi^2 - M^2) + Y (\phi^2 - M'^2) + Z( \chi \cdot \phi -
M''^2); \hspace{0.4in}
m_\chi^2 >0, \hspace{0.4in} m_\phi^2>0
\label{eq:Wlarge1}
\end{equation}
where we have also added a $Z$ driving term to fix the relative
orientation of $\vev{\chi}$ and $\vev{\phi}$. As before we may take $M$,
$M'$ and $M''$ real by a choice of the phases of the fields.
Minimizing the potential from $|F_X|^2$ the $SO(3)$ freedom allows the
choice: $\vev{\chi} = M (0,0,1)$. The minimization of $|F_Y|^2$ is not
identical, because now there is only a residual $SO(2)$ freedom, which
allows only the general form $\vev{\phi_i} = a_i + ib_i$, with $a_1 = 0$.
Setting $F_Y = 0$ and minimizing $\phi^* \phi$ gives $\vev{\phi} = M' (0,
\sin \theta, \cos \theta)$, with $\theta$ undetermined. The $Z$ driver
fixes $\cos \theta = M M'/ M''^2$ which is of order unity if $M, M'$ and
$M''$ are comparable. If all other flavons couple to the leptons through
higher dimension operators, $\theta_{atm} = \theta$.

Perhaps a more interesting case is to generate maximal mixing. To
achieve this, change $\vev{\phi}$ to the ``$U(1)$''
alignment of (\ref{eq:u1al})
\begin{equation}
W = X( \chi^2 - M^2) + Y \phi^2 + Z( \chi \cdot \phi - M''^2);
\hspace{0.4in} m_\chi^2 >0, \hspace{0.4in} m_\phi^2>0.
\label{eq:Wlarge2}
\end{equation}
As before, $SO(3)$ freedom allows $\vev{\chi} = M (0,0,1)$ and
$\vev{\phi_i} = a_i + ib_i$, with $a_1 = 0$. Setting $F_Z=0$ aligns $b_3
= 0$ and $a_3 = M''^2/M \equiv V$, while $F_Y = 0$ forces $b_1^2 + b_2^2 =
V^2 +
a_2^2$ and $a_2 b_2 = 0$. With $m_\phi^2 >0$, the remaining degeneracy
is completely lifted by the soft mass squared term, giving $a_2 = 0$ and
$(b_1,b_2) = V(\sin \theta, \cos \theta)$. Since $a_1 = a_2 = 0$, the
$SO(2)$ freedom has not been used up, and we can choose an $SO(2)$
basis in which $\theta =0$:
\begin{equation}
\vev{\chi} = M \pmatrix{ 0 \cr 0 \cr 1} \hspace{1in}
\vev{\phi} = V \pmatrix{ 0 \cr i \cr 1}.
\label{vevs45}
\end{equation}
As expected, these vevs show that $\chi$ has an ``$SO(2)$'' alignment, while
$\phi$ has a ``$U(1)$'' alignment. The alignment term ensures that $\phi$
and $\chi$ vevs are not orthogonal.
The lepton masses from (\ref{eq:massops}) now give $\theta_{atm} =
45^\circ$, up to corrections of relative order $m_\mu / m_\tau$.
In the $(1,-,+)$ basis, this theory has the leading terms of
texture C with
\begin{equation}
m_E = \pmatrix{0 & 0 & 0 \cr 0&\delta_2&0 \cr 0&0&D_3}\equiv m_{I}
\end{equation}

\section{Models With Large $\Omega_\nu$ and Large $\theta_{atm}$}

The key to avoiding the $\beta \beta_{0 \nu}$ constraint
(\ref{eq:doublebeta}) \cite{bimax,gg}
is to have a $U(1)$ vev alignment in the 12 space
so that the electron is in $l_-$. In this basis the $SO(3)$ invariant
neutrino mass term is $2 l_+ l_- + l_3 l_3$, as shown in texture B,
and gives a vanishing $\beta \beta_{0 \nu}$ rate.
Thus we seek to modify the model of eqn (\ref{eq:massops}), which
generates large $\theta_{atm}$, to align the electron along $l_-$.
The interactions of (\ref{eq:massops}) are
insufficient to identify the electron. We must add perturbations for the
muon mass, which will identify the electron as the massless state.
Hence we extend (\ref{eq:massops}) to
\begin{equation}
W_1 = (l \cdot l) hh + (l \cdot \chi)^2 hh + (l \cdot \phi_\tau) \tau h
+ (l \cdot \phi_\mu) \xi_\mu \mu h,
\label{eq:W1}
\end{equation}
and seek potentials where $\vev{\phi_{\tau, \mu}}$ have zero components
in the $+$ direction.

To obtain a $\chi$ vev in the 3 direction, and a $U(1)$ alignment in the
12 space, we use (\ref{eq:Wlarge2}), with $M'' = 0$ to enforce the
orthogonality of $\phi$ with $\chi$
\begin{equation}
W_2 = X( \chi^2 - M^2) + Y \phi_\mu^2 + Z \chi \cdot \phi_\mu;
\hspace{0.4in} m_\chi^2 >0, \hspace{0.4in} m_{\phi_\mu}^2 < 0,
\label{W2}
\end{equation}
which gives
\begin{equation}
\vev{\chi} = M \pmatrix{ 0 \cr 0 \cr 1} \hspace{1in}
\vev{\phi_\mu} = V \pmatrix{ 1 \cr i \cr 0}.
\label{vevchi}
\end{equation}
Large $\theta_{atm}$
requires $\vev{\phi_\tau}$ to have large components in
both $-$ and 3 directions, and results from the addition
\begin{equation}
W_3 = Z' \; \phi_\mu \cdot \phi_\tau;
\hspace{0.4in} m_{\phi_\tau}^2 < 0.
\label{W3}
\end{equation}
In the (1,2,3) basis
\begin{equation}
\vev{\phi_\tau} = V' \pmatrix{ 1 \cr i \cr \sqrt{2} x}.
\label{vevphitau}
\end{equation}
This theory, described by $W_1 + W_2 + W_3$,
determines the vev orientations so that $\Omega_\nu$ is
unconstrained by $\beta \beta_{0 \nu}$ decay. The value of
$\theta_{atm}$ is generically of order unity, but is not determined.

Additional potential terms can determine $x$ and hence $\theta_{atm}$.
For example,
maximal mixing can be obtained in a theory with three extra triplets,
$\phi_{1,2,3}$. Discrete symmetries are introduced so that none of these
fields couples to matter: the matter interactions remain those of
(\ref{eq:W1}). The field $\phi_1$ is driven to have an $SO(2)$
alignment, and also the product $\phi_1 \cdot \chi$ is driven to zero.
The $SO(2)$ freedom, not specified until now, then allows the form
$\vev{\phi_1} = V_1 (1,0,0)$. The other two triplets $\phi_2$ and
$\phi_3$ are driven just like $\phi_\mu$ and $\phi_{\tau}$
respectively: $\phi_2^2, \phi_2 \cdot \chi$ and $\phi_2 \cdot \phi_3$ are
all forced to vanish. However, the vevs are not identical to those of
$\phi_{\mu,\tau}$, because $\phi_2 \cdot \phi_\mu$ is forced to
be non-zero, so that the discrete $\pm$ choice of the $U(1)$ alignment is
opposite for $\phi_{2,3}$ compared with $\phi_{\mu,\tau}$:
\begin{equation}
\vev{\phi_2} = V_2 \pmatrix{ 1 \cr -i \cr 0} \hspace{1in}
\vev{\phi_3} = V_3 \pmatrix{ 1 \cr -i \cr \sqrt{2} y}.
\label{vevphi23}
\end{equation}
Maximal mixing follows from two further constraints: forcing $\phi_3
\cdot \phi_\tau$ to zero imposes $xy = -1$, while forcing
$\epsilon \phi_1 \phi_3 \phi_\tau$ to zero ($\epsilon$ is the tensor totally
antisymmetric in $SO(3)$ indices)
sets $y = -x$. Hence $(x,y) = (\pm 1, \mp 1)$, giving $\theta_{atm} =
45^\circ$.
The complete theory is described by $W_1 + W_2 + W_3 + W_4$, where
\begin{eqnarray}
W_4 &=& X_1(\phi_1^2 - M_1^2) + X_2 \; \phi_1 \cdot \chi + X_3 \phi_2^2
+ X_4 \; \phi_2 \cdot \chi + X_5 \; \phi_2 \cdot \phi_3 \nonumber \\
& & + X_6 \; (\phi_2 \cdot \phi_\mu-M_2^2) + X_7 \; \phi_3 \cdot \phi_\tau + X_8 \;
\epsilon \phi_1 \phi_3 \phi_\tau. 
\label{eq:W4}
\end{eqnarray}

There are other options for constructing theories with interesting
vacuum alignments. For example, doublets may be used as well as
triplets, and if $SO(3)$ is gauged, the aligning potential may arise
from $D$ terms as well as $F$ terms.

\section{Conclusions}

In this letter we made a counter-intuitive proposal for a theory of
lepton masses; in the limit of exact flavour symmetry, the three
neutrinos are massive and degenerate, while the three charged leptons
are massless. Such zeroth-order masses result when the three lepton
doublets form a real irreducible representation of some non-Abelian
flavour group --- for example, a triplet of $SO(3)$. A sequential
breaking of the flavour group then produces both a hierarchy of
charged lepton masses and a hierarchy of neutrino $\Delta m^2$. The
Majorana neutrino masses are small because, as always, they are second
order in weak symmetry breaking.

We showed that the $SO(3)$ symmetry breaking may follow a different path
in the charged and neutral sectors, leading to a vacuum misalignment
with interesting consequences. There can be large leptonic mixing
angles, with $45^\circ$ arising from the simplest misalignment
potentials. Such mixing can explain the atmospheric neutrino data while
allowing large amounts of neutrino hot dark matter. The latter is consistent
with the bounds on the $\beta \beta_{0 \nu}$ process because the symmetry
suppresses the Majorana mass matrix element $m_{\nu ee}$. Such a model can
give bimaximal mixing \cite{bimax,gg} with the large mixing angles very 
close to $45^\circ$.

\section*{Acknowledgements}

This work was supported in part by the
U.S. Department of Energy under Contracts DE-AC03-76SF00098, in part
by the National Science Foundation under grant PHY-95-14797. HM was
also supported by Alfred P. Sloan Foundation.

\end{document}